\documentstyle[12pt]{article}
\newcounter{eq}
\newcounter{sc}

\def\overleftrightarrow#1{\vbox{\ialign{##\crcr
 $\leftrightarrow$\crcr\noalign{\kern-1pt\nointerlineskip}
 $\hfil\displaystyle{#1}\hfil$\crcr}}}

\newlength{\minitwocolumn}
\begin{document}

%%%%%%%%%%%%%%%%%%%%%%%%%%%%%%%%%%%%%%%%%%%%%%%%%%%%%%%%%%%%%%%%%%
%%%%%%%%%%%%%%%%%%%%%%%% Title %%%%%%%%%%%%%%%%%%%%%%%%%%%%%%%%%%%
%%%%%%%%%%%%%%%%%%%%%%%%%%%%%%%%%%%%%%%%%%%%%%%%%%%%%%%%%%%%%%%%%%
\begin{flushright}
DPUR/TH/45\\
May, 2015\\
\end{flushright}
\vspace{20pt}

%\magnification=\magstep1
\pagestyle{empty}
\baselineskip15pt
%\font\cmssB=cmss17
%\font\cmssS=cmss10

\begin{center}
{\large\bf Conformal Higgs Gravity
\vskip 1mm }

\vspace{20mm}
Ichiro Oda \footnote{E-mail address:\ ioda@phys.u-ryukyu.ac.jp
}

\vspace{5mm}
           Department of Physics, Faculty of Science, University of the 
           Ryukyus,\\
           Nishihara, Okinawa 903-0213, Japan.\\

\end{center}

%\maketitle

\vspace{5mm}
\begin{abstract}
We construct a model of conformal gravity with Higgs field. 
This model has a positive Newton's constant and exhibits a novel symmetry 
breaking mechanism of gauge symmetries. A possible application to cosmology
is briefly mentioned. 
\end{abstract}

\newpage
\pagestyle{plain}
\pagenumbering{arabic}
%\setcounter{page}{1}

%%%%%%%%%%%%%%%%%%%%%%%%%%%%%%%%%%%%%%%%%%%%%%%%%%%%%%%%%%%%%%%%%%
%%%%%%%%%%%%%%%%%%%%%%%% Article %%%%%%%%%%%%%%%%%%%%%%%%%%%%%%%%%
%%%%%%%%%%%%%%%%%%%%%%%%%%%%%%%%%%%%%%%%%%%%%%%%%%%%%%%%%%%%%%%%%%

\rm
%%%%%%%%%%%%%%%%%%%%%%%%%%%%%%%%%%%%%%%%%%%%%%%%%%%%%%%%%%%%%%%%%%%%%
%%%%%%%%%%%%%%%%%%%%%%%%%%%%%%   SEC  1    %%%%%%%%%%%%%%%%%%%%%%%%%%
%%%%%%%%%%%%%%%%%%%%%%%%%%%%%%%%%%%%%%%%%%%%%%%%%%%%%%%%%%%%%%%%%%%%%
\section{Introduction}

Quantum gravity is a difficult subject with a host of conceptional and computational problems
which we are at present far from resolving. One of the difficult problems in quantum gravity is 
that the Einstein-Hilbert action is unrenormalizable so that it gives rise to intractable divergences
at loop levels. Of course, the simple reason behind nonrenormalizability is that the gravitational
constant has dimension of length squared, or equivalently that the gravitational field has canonical 
dimension of mass like gauge fields and therefore the Einstein-Hilbert term and the gravity-matter
interaction ones possess dimensions greater than four.

A way out of this difficulty is to modify the action by adding higher-derivative terms involving 
the curvature tensors \cite{Utiyama}. Since the higher-derivative terms are dominant at the short distance 
scales or in the high energy regime, this recipe fulfils the requirement of renormalizability \cite{Stelle}, 
but we have to pay the price of success. One serious difficulty of higher-derivative gravity is the occurrence
of massive ghosts which cause unitarity to fail \cite{Mario}. This is the main reason why this theory has thus far
received little attention. The other difficulty is more aesthetic than functional in the sense that 
when the higher-derivative terms are allowed to exist in an action, the number of the possible terms, 
which must transform as a scalar under general coordinate transformations, is in principle infinite. 

Whereas the former difficulty associated with the higher-derivative gravity is a very hard problem,
which we do not tackle in this article, the latter one could be resolved by appealing to a local
conformal symmetry, which is one reason behind motivations in the study at hand. Let us recall that
a conformal symmetry picks up a unique action, which takes the quadratic form of a conformal tensor,
among an infinite number of higher-derivative terms \cite{Weyl}.    

Another motivation of the present study comes from a recent important observation that the Standard Model 
seems to be valid all the way up to the Planck scale and there is no new physics between the electro-weak scale 
and the Planck one \cite{Froggatt}. Then, one interesting conjecture is that our universe has started 
at the Planck scale just at a very distinguished point \cite{Morozov} where the Higgs self-coupling 
and its beta function as well as its bare mass \cite{Kawai} are all vanishing, and various parameters 
such as the Higgs mass $125 GeV$ \cite{ATLAS, CMS} are fully determined by a fundamental theory defined 
at the Planck scale with a conformal symmetry, which protects the vacuum fluctuations of the Higgs field 
and solve the hierarchy problem \cite{Bardeen}. As a remakable scenario, the Higgs sector 
might naturally emerge as a Goldstone one associated with spontaneous symmetry breakdown of the conformal 
symmetry existing at the Planck scale, and this scenario could solve at the same time the hierarchy and 
the Landau pole problems \cite{Morozov}.

Being armed with these conjecture and scenario, it is natural to couple conformal gravity to the Standard Model 
and investigate whether the Higgs mechanism could be derived from conformal gravity or not. Incidentally,
in recent years, interest in conformal gravity has been revived since Mannheim and Kazanas obtained the spherically 
symmetric solution and pointed out that this solution might be useful in understanding the rotational curves
of galaxies without recourse to dark matter \cite{Kazanas}. \footnote{There are some objections on this issue 
\cite{Flanagan, Mannheim 1, Brihaye, Yoon}.}
However, even if the Mannheim's conformal gravity has some intriguing features, it has a manifect drawback
in that the theory is formulated only in the negative gravitational constant, so the gravitational interaction 
is not attractive but repulsive \cite{Mannheim 2, Mannheim 3}.  It seems to be difficult for this exotic feature 
of the negative gravitational constant to be reconciled with the observation of the cosmic microwave background. 
More recently, a gravity model with the positive gravitational constant has been proposed but the model breaks 
a conformal symmetry in an explicit manner \cite{Phillips}, so it is not a gravity model on the basis of 
the conformal symmetry. 
   
In the rest of this article, we will briefly elaborate on some aspects of the conformal gravity with the Higgs field 
and mention an application to cosmology.

%%%%%%%%%%%%%%%%%%%%%%%%%%%%%%%%%%%%%%%%%%%%%%%%%%%%%%%%%%%%%%%%%%%%%
%%%%%%%%%%%%%%%%%%%%%%%%%%%%%%   SEC  2    %%%%%%%%%%%%%%%%%%%%%%%%%%
%%%%%%%%%%%%%%%%%%%%%%%%%%%%%%%%%%%%%%%%%%%%%%%%%%%%%%%%%%%%%%%%%%%%%
\section{A conformal gravity with the Higgs field}

We will start with an abelian model where a  $U(1)$ gauge field $A_\mu$ couples to conformal gravity
with two scalar fields, one of which, $\phi$, is real while the other, $H$, which is nothing but the
Higgs doublet field, is complex. In this article, we will ignore fermions completely since their existence 
does not change our conclusion at all. Moreover, the generalization to the non-abelian gauge fields
is straightforward. In other words, although the model proposed in this section is a toy model
describing some features of the Standard Model, it is straightforward to construct a more realistic 
model such that conformal gravity coexists with the Standard Model by following the same line of the
arugument as in this toy model.

With a background curved metric $g_{\mu\nu}$, our action is defined as follows: \footnote{We follow 
notation and conventions by Misner et al.'s textbook \cite{MTW}, for instance, the flat Minkowski metric
$\eta_{\mu\nu} = diag(-, +, +, +)$, the Riemann curvature tensor $R^\mu \ _{\nu\alpha\beta} = 
\partial_\alpha \Gamma^\mu_{\nu\beta} - \partial_\beta \Gamma^\mu_{\nu\alpha} + \Gamma^\mu_{\sigma\alpha} 
\Gamma^\sigma_{\nu\beta} - \Gamma^\mu_{\sigma\beta} \Gamma^\sigma_{\nu\alpha}$, 
and the Ricci tensor $R_{\mu\nu} = R^\alpha \ _{\mu\alpha\nu}$.
The reduced Planck mass is defined as $M_p = \sqrt{\frac{c \hbar}{8 \pi G}} = 2.4 \times 10^{18} GeV$.
Through this article, we adopt the reduced Planck units where we set $c = \hbar = M_p = 1$. 
In this units, all quantities become dimensionless. 
Finally, note that in the reduced Planck units, the Einstein-Hilbert Lagrangian density takes the form
${\cal L}_{EH} = \frac{1}{2} \sqrt{-g} R$.}
%**   Action 1  %%%%%%%%%%%%%%%%%%%%%%%%%%%%%%%%%%%%%%%%%%%%%%%%%%%%%%%%%
\begin{eqnarray}
S = \int d^4 x {\cal L} = S_W + S_M = \int d^4 x \left( {\cal L}_W + {\cal L}_M \right).
\label{Action 1}
\end{eqnarray}
%%%%%%%%%%%%%%%%%%%%%%%%%%%%%%%%%%%%%%%%%%%%%%%%%%%%%%%%%%%%%%%%%%%
Here the Lagrangian of conformal gravity takes the well-known form
%**   Conf-Action 1  %%%%%%%%%%%%%%%%%%%%%%%%%%%%%%%%%%%%%%%%%%%%%%%%%%%%%%%%%
\begin{eqnarray}
\frac{1}{\sqrt{-g}}{\cal L}_W &=&  - \alpha_g C_{\mu\nu\rho\sigma} C^{\mu\nu\rho\sigma}
\nonumber\\
&=& - \alpha_g \left( R_{\mu\nu\rho\sigma} R^{\mu\nu\rho\sigma} - 2 R_{\mu\nu} R^{\mu\nu}
+ \frac{1}{3} R^2 \right)
\nonumber\\
&=& - 2 \alpha_g \left( R_{\mu\nu} R^{\mu\nu} - \frac{1}{3} R^2 \right),
\label{Conf-Action 1}
\end{eqnarray}
%%%%%%%%%%%%%%%%%%%%%%%%%%%%%%%%%%%%%%%%%%%%%%%%%%%%%%%%%%%%%%%%%%%
where in the last equality we have used the fact that 
$\sqrt{-g} \left( R_{\mu\nu\rho\sigma} R^{\mu\nu\rho\sigma} - 4 R_{\mu\nu} R^{\mu\nu} + R^2 \right)$
is a total divergence. The coefficient $\alpha_g$ is a dimensionless constant reflecting
a conformal symmetry. The conformal tensor $C_{\mu\nu\rho\sigma}$, which is sometimes called 
the Weyl tensor as well, is defined as
%**   Conf-Tensor  %%%%%%%%%%%%%%%%%%%%%%%%%%%%%%%%%%%%%%%%%%%%%%%%%%%%%%%%%
\begin{eqnarray}
C_{\mu\nu\rho\sigma} = R_{\mu\nu\rho\sigma} - \left( g_{\mu [\rho} R_{\sigma] \nu}
- g_{\nu [\rho} R_{\sigma] \mu} \right) + \frac{1}{3} g_{\mu [\rho} g_{\sigma] \nu} R, 
\label{Conf-Tensor}
\end{eqnarray}
%%%%%%%%%%%%%%%%%%%%%%%%%%%%%%%%%%%%%%%%%%%%%%%%%%%%%%%%%%%%%%%%%%%
where we have introduced a notation, $A_{[\mu} B_{\nu]} = \frac{1}{2} (A_\mu B_\nu - A_\nu B_\mu)$.
The conformal tensor shares the same symmetric properties of indices as those of the Riemann tensor
and is in addition trace-free on all its indices. It also behaves in a very simple way under conformal 
transformation of the metric, $g_{\mu\nu} \rightarrow \Omega^2 (x) g_{\mu\nu}$, 
like $C^\mu \ _{\nu\rho\sigma} \rightarrow C^\mu \ _{\nu\rho\sigma}$.   

The Lagrangian of the matter fields takes the form \footnote{Similar Lagrangians are manipulated
in different contexts \cite{Oda1}-\cite{Bars2}.}
%**   Matt-Action 1  %%%%%%%%%%%%%%%%%%%%%%%%%%%%%%%%%%%%%%%%%%%%%%%%%%%%%%%%%
\begin{eqnarray}
\frac{1}{\sqrt{-g}}{\cal L}_M &=&  \frac{1}{12} \left( \phi^2 - 2 |H|^2 \right) R + \frac{1}{2} g^{\mu\nu} 
\partial_\mu \phi \partial_\nu \phi - g^{\mu\nu} (D_\mu H)^\dagger D_\nu H    \nonumber\\
&-& \frac{1}{4} g^{\mu\nu} g^{\rho\sigma} F_{\mu\rho} F_{\nu\sigma} - V(H, \phi),
\label{Matt-Action 1}
\end{eqnarray}
%%%%%%%%%%%%%%%%%%%%%%%%%%%%%%%%%%%%%%%%%%%%%%%%%%%%%%%%%%%%%%%%%%%
where $\phi$ is essentially a ghost, but this is not a problem because if necessary it can be removed by fixing 
the local conformal symmetry. The covariant derivative and field strength in the abelian gauge group are 
respectively defined as 
%**   Def 1  %%%%%%%%%%%%%%%%%%%%%%%%%%%%%%%%%%%%%%%%%%%%%%%%%%%%%%%%%
\begin{eqnarray}
D_\mu H = (\partial_\mu + i g A_\mu) H,   \quad
(D_\mu H)^\dagger = H^\dagger (\overleftarrow{\partial}_\mu - i g A_\mu) ,   \quad
F_{\mu\nu} = \partial_\mu A_\nu - \partial_\nu A_\mu,
\label{Def 1}
\end{eqnarray}
%%%%%%%%%%%%%%%%%%%%%%%%%%%%%%%%%%%%%%%%%%%%%%%%%%%%%%%%%%%%%%%%%%%
with $g$ being a $U(1)$ real coupling constant. (We use the same alphabet "$g$" to denote the gauge coupling
and the determinant of the metric tensor, but the difference would be obvious from the context 
since the latter always appears in the form of $\sqrt{-g}$.)

The action (\ref{Action 1}) is invariant under a local conformal transformation (or Weyl transformation). 
In fact, with a local parameter $\Omega(x)$, which is a finite, non-vanishing and continuous real function, 
the conformal transformation is defined as 
%**   Weyl transf %%%%%%%%%%%%%%%%%%%%%%%%%%%%%%%%%%%%%%%%%%%%%%%%%%%%%%%%%
\begin{eqnarray}
g_{\mu\nu} &\rightarrow& \tilde g_{\mu\nu} = \Omega^2(x) g_{\mu\nu},  \quad
g^{\mu\nu} \rightarrow \tilde g^{\mu\nu} = \Omega^{-2}(x) g^{\mu\nu}, \quad
 \nonumber\\
\phi &\rightarrow& \tilde \phi = \Omega^{-1}(x) \phi, \quad
H \rightarrow \tilde H = \Omega^{-1}(x) H,  \quad
A_\mu \rightarrow \tilde A_\mu = A_\mu.
\label{Weyl transf}
\end{eqnarray}
%%%%%%%%%%%%%%%%%%%%%%%%%%%%%%%%%%%%%%%%%%%%%%%%%%%%%%%%%%%%%%%%%%%

In this article, we will fully make use of the unitary gauge $H(x) = \frac{1}{\sqrt{2}} e^{i \alpha \theta(x)} 
(0, h(x))^T$ where $\alpha$, $\theta(x)$ and $h(x)$ are respectively a real number, the Nambu-Goldstone boson 
and the physical Higgs field. With the unitary gauge, the Lagrangian of the matter fields (\ref{Matt-Action 1})
can be rewritten as
%**   Matt-Action 2  %%%%%%%%%%%%%%%%%%%%%%%%%%%%%%%%%%%%%%%%%%%%%%%%%%%%%%%%%
\begin{eqnarray}
\frac{1}{\sqrt{-g}}{\cal L}_M =  \frac{1}{12} \left( \phi^2 - h^2 \right) R  
+ \frac{1}{2} (\partial_\mu \phi)^2 - \frac{1}{2} \left[ (\partial_\mu h)^2 
+ g^2 B_\mu^2 h^2 \right] - \frac{1}{4} F_{\mu\nu}^2 - V(h, \phi),
\label{Matt-Action 2}
\end{eqnarray}
%%%%%%%%%%%%%%%%%%%%%%%%%%%%%%%%%%%%%%%%%%%%%%%%%%%%%%%%%%%%%%%%%%%
where we have defined a new gauge field $B_\mu$ as $B_\mu = A_\mu + \frac{\alpha}{g} \partial_\mu \theta$
and consequently the gauge strength is now given by $F_{\mu\nu} \equiv \partial_\mu A_\nu - \partial_\nu A_\mu
= \partial_\mu B_\nu - \partial_\nu B_\mu$. 
In order to keep the potential $V(h, \phi)$ be invariant under the conformal transformation 
(\ref{Weyl transf}), the potential $V(h, \phi)$ must be a polynomial of the fourth degree 
in $h$ and $\phi$ 
%**   Potential 1  %%%%%%%%%%%%%%%%%%%%%%%%%%%%%%%%%%%%%%%%%%%%%%%%%%%%%%%%%
\begin{eqnarray}
V(h, \phi) =  a_1 h^4 + a_2 h^2 \phi^2 + a_3 \phi^4,
\label{Potential 1}
\end{eqnarray}
%%%%%%%%%%%%%%%%%%%%%%%%%%%%%%%%%%%%%%%%%%%%%%%%%%%%%%%%%%%%%%%%%%%
where $a_1, a_2, a_3$ are constants. \footnote{In the most general case, these coefficients can be functions 
of the dimensionless variable $z \equiv \frac{\sqrt{2 |H|^2}}{\phi} = \frac{h}{\phi}$, but for simplicity
we shall assume them to be constants \cite{Oda4}.} 

It is straightforward to derive the field equations with respect to the metric tensor $g_{\mu\nu}$.
First, variation of the part of conformal gravity reads
%**   Var-Cof-Grav  %%%%%%%%%%%%%%%%%%%%%%%%%%%%%%%%%%%%%%%%%%%%%%%%%%%%%%%%%
\begin{eqnarray}
\frac{2}{\sqrt{-g}} \frac{\delta S_W}{\delta g_{\mu\nu}} =  - 4 \alpha_g W^{\mu\nu}.
\label{Var-Cof-Grav}
\end{eqnarray}
%%%%%%%%%%%%%%%%%%%%%%%%%%%%%%%%%%%%%%%%%%%%%%%%%%%%%%%%%%%%%%%%%%%
Here we have defined $W^{\mu\nu} \equiv W^{\mu\nu}_{(2)} - \frac{1}{3} W^{\mu\nu}_{(1)}$
where $W^{\mu\nu}_{(1)}$ and $W^{\mu\nu}_{(2)}$ are respectively defined as
%**   W1  %%%%%%%%%%%%%%%%%%%%%%%%%%%%%%%%%%%%%%%%%%%%%%%%%%%%%%%%%
\begin{eqnarray}
W^{\mu\nu}_{(1)} &\equiv& \frac{1}{\sqrt{-g}} \frac{\delta}{\delta g_{\mu\nu}} \int d^4 x \sqrt{-g} R^2
\nonumber\\
&=& - 2 g^{\mu\nu} \nabla^2 R + 2 \nabla^\mu \nabla^\nu R - 2 R R^{\mu\nu} + \frac{1}{2} g^{\mu\nu} R^2.
\label{W1}
\end{eqnarray}
%%%%%%%%%%%%%%%%%%%%%%%%%%%%%%%%%%%%%%%%%%%%%%%%%%%%%%%%%%%%%%%%%%%
%**   W2  %%%%%%%%%%%%%%%%%%%%%%%%%%%%%%%%%%%%%%%%%%%%%%%%%%%%%%%%%
\begin{eqnarray}
W^{\mu\nu}_{(2)} &\equiv& \frac{1}{\sqrt{-g}} \frac{\delta}{\delta g_{\mu\nu}} \int d^4 x \sqrt{-g} 
R_{\mu\nu} R^{\mu\nu}
\nonumber\\
&=& - \frac{1}{2} g^{\mu\nu} \nabla^2 R - \nabla^2 R^{\mu\nu} + \nabla_\rho \nabla^\nu R^{\mu\rho}
+ \nabla_\rho \nabla^\mu R^{\nu\rho} - 2 R^{\mu\rho} R^\nu \ _\rho + \frac{1}{2} g^{\mu\nu} R_{\rho\sigma}^2.
\label{W2}
\end{eqnarray}
%%%%%%%%%%%%%%%%%%%%%%%%%%%%%%%%%%%%%%%%%%%%%%%%%%%%%%%%%%%%%%%%%%%
Next, for variation of the matter sector, we have the energy-momentum tensor $T^{\mu\nu}$
%**   Var-Matt  %%%%%%%%%%%%%%%%%%%%%%%%%%%%%%%%%%%%%%%%%%%%%%%%%%%%%%%%%
\begin{eqnarray}
\frac{2}{\sqrt{-g}} \frac{\delta S_M}{\delta g_{\mu\nu}} =  T^{\mu\nu}.
\label{Var-Matt}
\end{eqnarray}
%%%%%%%%%%%%%%%%%%%%%%%%%%%%%%%%%%%%%%%%%%%%%%%%%%%%%%%%%%%%%%%%%%%
Here $T_{\mu\nu}$ is of form
%**   T-1  %%%%%%%%%%%%%%%%%%%%%%%%%%%%%%%%%%%%%%%%%%%%%%%%%%%%%%%%%
\begin{eqnarray}
T_{\mu\nu} &=&  - \frac{1}{6} \left( \phi^2 - h^2 \right) \left( R_{\mu\nu} - \frac{1}{2} g_{\mu\nu} R \right)
- \frac{2}{3} \partial_\mu \phi \partial_\nu \phi + \frac{1}{6} g_{\mu\nu} (\partial_\rho \phi)^2
\nonumber\\
&+& \frac{2}{3} \partial_\mu h \partial_\nu h - \frac{1}{6} g_{\mu\nu} (\partial_\rho h)^2
+ g^2 \left( B_\mu B_\nu - \frac{1}{2} g_{\mu\nu} B_\rho ^2 \right) h^2
\nonumber\\
&+& \frac{1}{3} \left[ \phi \nabla_\mu \nabla_\nu \phi - h \nabla_\mu \nabla_\nu h
- g_{\mu\nu} \phi \nabla^\rho \nabla_\rho \phi + g_{\mu\nu} h \nabla^\rho \nabla_\rho h \right]
\nonumber\\
&+& F_{\mu\rho} F_\nu \ ^\rho - \frac{1}{4} g_{\mu\nu} F_{\rho\sigma}^2 - g_{\mu\nu} V(h, \phi).
\label{T-1}
\end{eqnarray}
%%%%%%%%%%%%%%%%%%%%%%%%%%%%%%%%%%%%%%%%%%%%%%%%%%%%%%%%%%%%%%%%%%%
Thus,  the field equations for the metric reads \cite{Mannheim 2, Mannheim 3}
%**   Eq-g  %%%%%%%%%%%%%%%%%%%%%%%%%%%%%%%%%%%%%%%%%%%%%%%%%%%%%%%%%
\begin{eqnarray}
4 \alpha_g W^{\mu\nu} = T^{\mu\nu}.
\label{Eq-g}
\end{eqnarray}
%%%%%%%%%%%%%%%%%%%%%%%%%%%%%%%%%%%%%%%%%%%%%%%%%%%%%%%%%%%%%%%%%%%

Notice that the both sides in Eq. (\ref{Eq-g}) should be traceless owing to conformal invariance.
Actually, the tensor $W^{\mu\nu}$ is traceless because of the Bianchi identity. To show that
$T^{\mu\nu}$ is traceless, one has to use the field equations for $\phi$ and $h$, which are 
given by
%**   Eq-phi & h  %%%%%%%%%%%%%%%%%%%%%%%%%%%%%%%%%%%%%%%%%%%%%%%%%%%%%%%%%
\begin{eqnarray}
&{}& \frac{1}{6} \phi R - \nabla^2 \phi - \frac{\partial V}{\partial \phi} = 0,
\nonumber\\
&{}& \frac{1}{6} h R - \nabla^2 h + g^2 B_\mu ^2 h + \frac{\partial V}{\partial h} = 0.
\label{Eq-phi & h}
\end{eqnarray}
%%%%%%%%%%%%%%%%%%%%%%%%%%%%%%%%%%%%%%%%%%%%%%%%%%%%%%%%%%%%%%%%%%%
Using these field equations, the trace part of $T^{\mu\nu}$ takes the form
%**   Trace  %%%%%%%%%%%%%%%%%%%%%%%%%%%%%%%%%%%%%%%%%%%%%%%%%%%%%%%%%
\begin{eqnarray}
T^\mu _\mu = \phi \frac{\partial V}{\partial \phi} + h \frac{\partial V}{\partial h}
- 4 V.
\label{Trace}
\end{eqnarray}
%%%%%%%%%%%%%%%%%%%%%%%%%%%%%%%%%%%%%%%%%%%%%%%%%%%%%%%%%%%%%%%%%%%
This quantity becomes vanishing when the potential $V$ is a polynomial of the fourth degree
in $h$ and $\phi$ as in Eq. (\ref{Potential 1}).

%%%%%%%%%%%%%%%%%%%%%%%%%%%%%%%%%%%%%%%%%%%%%%%%%%%%%%%%%%%%%%%%%%%%%
%%%%%%%%%%%%%%%%%%%%%%%%%%%%%%   SEC  3    %%%%%%%%%%%%%%%%%%%%%%%%%%
%%%%%%%%%%%%%%%%%%%%%%%%%%%%%%%%%%%%%%%%%%%%%%%%%%%%%%%%%%%%%%%%%%%%%
\section{Einstein gauge for conformal symmetry}

In the previous section, we have constructed a model of conformal gravity coupled to 
the ghost-like scalar field, the Higgs field and the abelian gauge field. 
Based on this model, we will construct a model of conformal gravity with the positive Newton's constant.

Before doing so, it is worthwhile to consider why the Mannheim's model of conformal gravity has
the negative effective gravitational constant \cite{Mannheim 2, Mannheim 3}. 
In his model, there is only a ghost-like scalar field 
$\phi$ but no Higgs field $h$. In order to avoid the ghost, Mannheim has changed the overall
sign of the action of the scalar sector, and then taken a gauge condition $\phi = constant$
for a local conformal symmetry. This procedure naturally leads to the negative effective Newton's
constant. To put differently, the ghost-like scalar $\phi$ produces the negative gravitational
constant via its conformal coupling to gravity if it is interpreted as a normal scalar. 
  
On the other hand, in our model, in addition to the ghost-like scalar $\phi$ there is the Higgs field
$h$ with the healthy kinetic term, so there is a possibility of obtaining the positive gravitational
constant.  

To do that, let us take the "Einstein gauge (E-gauge)" for a local conformal transformation 
(For clarity, we will recover the Planck mass in the following two equations.)
%**   E-gauge %%%%%%%%%%%%%%%%%%%%%%%%%%%%%%%%%%%%%%%%%%%%%%%%%%%%%%%%%
\begin{eqnarray}
\phi^2 - h^2 = 6 M_p^2.      
\label{E-gauge}
\end{eqnarray}
%%%%%%%%%%%%%%%%%%%%%%%%%%%%%%%%%%%%%%%%%%%%%%%%%%%%%%%%%%%%%%%%%%%
This $SO(1,1)$ invariant gauge choice can be parametrized in terms of a canonically normalized
real scalar field $\varphi$ as
%**   Parametrization  %%%%%%%%%%%%%%%%%%%%%%%%%%%%%%%%%%%%%%%%%%%%%%%%%%%%%%%%%
\begin{eqnarray}
\phi = \sqrt{6} M_p \cosh \frac{\varphi}{\sqrt{6} M_p},   \quad
h = \sqrt{6} M_p \sinh \frac{\varphi}{\sqrt{6} M_p}.
\label{Parametrization}
\end{eqnarray}
%%%%%%%%%%%%%%%%%%%%%%%%%%%%%%%%%%%%%%%%%%%%%%%%%%%%%%%%%%%%%%%%%%%
With the E-gauge (and the unitary gauge), the matter Lagrangian (\ref{Matt-Action 2}) can be 
cast to
%**   Matt-Action 3  %%%%%%%%%%%%%%%%%%%%%%%%%%%%%%%%%%%%%%%%%%%%%%%%%%%%%%%%%
\begin{eqnarray}
\frac{1}{\sqrt{-g}}{\cal L}_M =  \frac{1}{2} R - \frac{1}{2}  (\partial_\mu \varphi)^2 
- 3 g^2 B_\mu ^2 \sinh^2 \frac{\varphi}{\sqrt{6}} - \frac{1}{4} F_{\mu\nu}^2  - V(h, \phi).
\label{Matt-Action 3}
\end{eqnarray}
%%%%%%%%%%%%%%%%%%%%%%%%%%%%%%%%%%%%%%%%%%%%%%%%%%%%%%%%%%%%%%%%%%%
Note that the first term is nothing but the conventional Einstein-Hilbert term. We have
therefore derived the positive Newton's constant in the E-gauge by starting with 
the conformally-invariant action (\ref{Action 1}).

At this stage, let us derive the energy-momentum tensor in the E-gauge
%**   T-2  %%%%%%%%%%%%%%%%%%%%%%%%%%%%%%%%%%%%%%%%%%%%%%%%%%%%%%%%%
\begin{eqnarray}
T_{\mu\nu} &=&  - \left( R_{\mu\nu} - \frac{1}{2} g_{\mu\nu} R \right)
+ \partial_\mu \varphi \partial_\nu \varphi - \frac{1}{2} g_{\mu\nu} (\partial_\rho \varphi)^2
\nonumber\\
&+& 6 g^2 \left( B_\mu B_\nu - \frac{1}{2} g_{\mu\nu} B_\rho ^2 \right) \sinh^2 \frac{\varphi}{\sqrt{6}}
+ F_{\mu\rho} F_\nu \ ^\rho - \frac{1}{4} g_{\mu\nu} F_{\rho\sigma}^2 - g_{\mu\nu} V(h, \phi).
\label{T-2}
\end{eqnarray}
%%%%%%%%%%%%%%%%%%%%%%%%%%%%%%%%%%%%%%%%%%%%%%%%%%%%%%%%%%%%%%%%%%%
Since the equations of motion for the metric tensor have the same form as Eq. (\ref{Eq-g}) and the tensor
$W^{\mu\nu}$ is traceless, the traceless condition of $T_{\mu\nu}$ comes from the equations of motion 
for the metric tensor even if we have fixed the conformal symmetry in the E-gauge. In other words,
in this model, $T_\mu^\mu = 0$ is a part of the field equations. This point is one of interesting features
in the model at hand.

Now we wish to specify the form of the potential $V(h, \phi)$. Since we bear conformal gravity coupled to
the Standard Model in mind, it is natural to require that the potential $V(h, \phi)$ in Eq. (\ref{Potential 1}) 
should take the form of the usual Higgs potential 
%**   Potential 2  %%%%%%%%%%%%%%%%%%%%%%%%%%%%%%%%%%%%%%%%%%%%%%%%%%%%%%%%%
\begin{eqnarray}
V(h, \phi) &=&  a_1 h^4 + a_2 h^2 \phi^2 + a_3 \phi^4,
\nonumber\\
&=& \frac{\lambda}{8} ( h^2 - v^2 )^2 + \Lambda,
\label{Potential 2}
\end{eqnarray}
%%%%%%%%%%%%%%%%%%%%%%%%%%%%%%%%%%%%%%%%%%%%%%%%%%%%%%%%%%%%%%%%%%%
where $v$ and $\Lambda$ are constants.
Here the constants $a_1, a_2, a_3$ are chosen to be
%**   Constants  %%%%%%%%%%%%%%%%%%%%%%%%%%%%%%%%%%%%%%%%%%%%%%%%%%%%%%%%
\begin{eqnarray}
a_1 &=&  \frac{1}{36} \Lambda + \frac{\lambda}{8} \left( 1 + \frac{v^2}{6} \right)^2,  \quad
a_2 =  - \frac{1}{18} \Lambda - \frac{\lambda}{8} \left( \frac{v^4}{18} + \frac{v^2}{3} \right),
\nonumber\\
a_3 &=&  \frac{1}{36} \Lambda + \frac{\lambda}{8} \frac{v^4}{36}.
\label{Constants}
\end{eqnarray}
%%%%%%%%%%%%%%%%%%%%%%%%%%%%%%%%%%%%%%%%%%%%%%%%%%%%%%%%%%%%%%%%%%%
Here it is worth pointing out one important difference between the potential $V(h, \phi)$ in Eq. (\ref{Potential 2})
and the conventional Higgs potential though an appearance is similar. Recall that we have started with 
a completely conformally-invariant model, then fixed the local conformal symmetry by the $SO(1,1)$ invariant gauge
condition (\ref{E-gauge}) and consequently we have arrived at a potential of the form $V = c_1 h^4 + c_2 h^2 + c_3$
with $c_1, c_2, c_3$ are some constants expressed in terms of the original constants $a_1, a_2, a_3$.
Thus, in our model, the gauge fixing of the conformal symmetry gives rise to a potential of the Higgs type.
This fact should be contrasted to the situation in the Standard Model where the existence of the potential
$V = c_1 h^4 + c_2 h^2 + c_3$ is assumed in an ad hoc manner. 

Next let us note that for $\varphi \ll M_p$, or equivalently in the low energy regime, via Eq. (\ref{Parametrization})
we can approximate the Higgs field by the scalar field $\varphi$ like $h \approx \varphi$. Expanding $\varphi$
around $v$ like $\varphi = v + \hat \varphi$, the total Lagrangian of conformal gravity plus the matter fields 
is reduced to
%**   Total Action  %%%%%%%%%%%%%%%%%%%%%%%%%%%%%%%%%%%%%%%%%%%%%%%%%%%%%%%%%
\begin{eqnarray}
\frac{1}{\sqrt{-g}}{\cal L} &=&  - \alpha_g C_{\mu\nu\rho\sigma}^2 + \frac{1}{2} R 
- \frac{1}{2}  (\partial_\mu \hat \varphi)^2 - \frac{1}{2} \lambda v^2 \hat \varphi^2
- \frac{1}{4} F_{\mu\nu}^2 
\nonumber\\ 
&-& \frac{1}{2} g^2 v^2 B_\mu^2 - \frac{1}{2} g^2 B_\mu^2 ( 2 v \hat \varphi + \hat \varphi^2 )
- \frac{\lambda}{8} ( 4 v \hat \varphi^3 + \hat \varphi^4 ).
\label{Total Action}
\end{eqnarray}
%%%%%%%%%%%%%%%%%%%%%%%%%%%%%%%%%%%%%%%%%%%%%%%%%%%%%%%%%%%%%%%%%%%
This Lagrangian precisely describes a model where conformal gravity and the Einstein one 
couple to the physical Higgs field $\hat \varphi$ of mass $\sqrt{\lambda} v$ and the abelian gauge
field $B_\mu$ of mass $gv$ in addition to the conventional interactions between the Higgs field and
the gauge one. Accordingly, starting with a completely conformal invariant theory, we have not only succeeded in 
getting a conformal gravity with the positive Newton's constant but also in getting a toy model 
closely describing the Standard Model coupling with the conformal gravity plus the Einstein gravity. \footnote
{Indeed, it is straightforward to extend the matter sector to the Standard Model.}

%%%%%%%%%%%%%%%%%%%%%%%%%%%%%%%%%%%%%%%%%%%%%%%%%%%%%%%%%%%%%%%%%%%%%
%%%%%%%%%%%%%%%%%%%%%%%%%%%%%%   SEC  4    %%%%%%%%%%%%%%%%%%%%%%%%%%
%%%%%%%%%%%%%%%%%%%%%%%%%%%%%%%%%%%%%%%%%%%%%%%%%%%%%%%%%%%%%%%%%%%%%
\section{An application to cosmology}

Now we are ready to present the simplest application to cosmology.
In this section, we will consider the Robertson-Walker cosmological model whose line element is of form
%**   RW  %%%%%%%%%%%%%%%%%%%%%%%%%%%%%%%%%%%%%%%%%%%%%%%%%%%%%%%%%
\begin{eqnarray}
d s^2 &=& g_{\mu\nu} d x^\mu d x^\nu 
\nonumber\\ 
&=& - d t^2 + a^2(t) \left( \frac{1}{1 - kr^2} d r^2 + r^2 d \Omega^2 \right),
\label{RW}
\end{eqnarray}
%%%%%%%%%%%%%%%%%%%%%%%%%%%%%%%%%%%%%%%%%%%%%%%%%%%%%%%%%%%%%%%%%%%
where $d \Omega^2$ is the shorthand for $d \theta^2 + \sin^2 \theta d \phi^2$.
Closed, open and flat universes correspond to $k = 1, -1, 0$, respectively.
For closed and open universes, the scale factor $a(t)$ is the radius of spatial
curvature at given moment of time, while for spatially flat universe, the scale 
factor itself does not have any physical meaning since it can be set to any number
by rescaling the spatial coordinates. What is physically significant in this section 
is the Hubble parameter $H (t)$ and the dimensionless parameter $\Omega_\Lambda (t)$ representing 
a relative contribution of the cosmological constant to the density parameter, which
are defined by
%**   Hubble  %%%%%%%%%%%%%%%%%%%%%%%%%%%%%%%%%%%%%%%%%%%%%%%%%%%%%
\begin{eqnarray}
H (t) = \frac{\dot a (t)}{a (t)}, \quad \Omega_\Lambda (t) = \frac{\Lambda}{3 H^2 (t)}.
\label{Hubble-p}
\end{eqnarray}
%%%%%%%%%%%%%%%%%%%%%%%%%%%%%%%%%%%%%%%%%%%%%%%%%%%%%%%%%%%%%%%%%%%

In this article, we will confine ourselves to a classical analysis and neglect quantum corrections.
Then, since the Robertson-Walker metric (\ref{RW}) is known to be conformally flat, the tensor $W_{\mu\nu}$ 
becomes vanishing, thereby reducing the field equations (\ref{Eq-g}) to a simple expression
%**   Eq-g2  %%%%%%%%%%%%%%%%%%%%%%%%%%%%%%%%%%%%%%%%%%%%%%%%%%%%%%%%%
\begin{eqnarray}
T_{\mu\nu} = 0.
\label{Eq-g2}
\end{eqnarray}
%%%%%%%%%%%%%%%%%%%%%%%%%%%%%%%%%%%%%%%%%%%%%%%%%%%%%%%%%%%%%%%%%%%
Together with Eq. (\ref{T-2}) and Eq. (\ref{Potential 2}), Eqs. (\ref{Eq-g2}) produce the Einstein equations 
%**   Einstein 1  %%%%%%%%%%%%%%%%%%%%%%%%%%%%%%%%%%%%%%%%%%%%%%%%%%%%%%%%%
\begin{eqnarray}
R_{\mu\nu} - \frac{1}{2} g_{\mu\nu} R + \Lambda g_{\mu\nu} = \tilde T_{\mu\nu},
\label{Einstein 1}
\end{eqnarray}
%%%%%%%%%%%%%%%%%%%%%%%%%%%%%%%%%%%%%%%%%%%%%%%%%%%%%%%%%%%%%%%%%%%
where the newly-defined energy-momentum tensor $\tilde T_{\mu\nu}$ is given by
%**   Einstein 2  %%%%%%%%%%%%%%%%%%%%%%%%%%%%%%%%%%%%%%%%%%%%%%%%%%%%%%%%%
\begin{eqnarray}
\tilde T_{\mu\nu} &=& \partial_\mu \varphi \partial_\nu \varphi - \frac{1}{2} g_{\mu\nu} (\partial_\rho \varphi)^2
+ 6 g^2 \left( B_\mu B_\nu - \frac{1}{2} g_{\mu\nu} B_\rho ^2 \right) \sinh^2 \frac{\varphi}{\sqrt{6}}
\nonumber\\
&+& F_{\mu\rho} F_\nu \ ^\rho - \frac{1}{4} g_{\mu\nu} F_{\rho\sigma}^2 
- \frac{\lambda}{8} g_{\mu\nu} \left( 6 \sinh^2 \frac{\varphi}{\sqrt{6}} - v^2 \right)^2.
\label{Einstein 2}
\end{eqnarray}
%%%%%%%%%%%%%%%%%%%%%%%%%%%%%%%%%%%%%%%%%%%%%%%%%%%%%%%%%%%%%%%%%%%

At the lowest level of the approximation $< \varphi > \approx < h > = v$, the trace part of this
energy-momentum tensor reads
%**   Trace part %%%%%%%%%%%%%%%%%%%%%%%%%%%%%%%%%%%%%%%%%%%%%%%%%%%%%%%%%
\begin{eqnarray}
\tilde T^\mu_\mu = - g^2 v^2 B_\mu ^2.
\label{Trace part}
\end{eqnarray}
%%%%%%%%%%%%%%%%%%%%%%%%%%%%%%%%%%%%%%%%%%%%%%%%%%%%%%%%%%%%%%%%%%%
Thus, in the energy region $M_{EM} < E < M_p$ where $M_{EM}$ is the electro-weak energy scale,
one could regard the energy-momentum tensor as $\tilde T^\mu_\mu \approx 0$. This approximation
is physically plausible since all particles would be massless well above the electro-weak scale.
To put differently, in applying the model at hand to cosmology, we will consider only the epoch
where our universe starts at the Planck scale and then cool down before the electro-weak phase transition.
In this epoch, a conformal symmetry is unbroken so that the masses of vector bosons and fermions are not
generated.  
 
Following the standard technique of a cosmological approach, we express the energy-momentum tensor
in terms of a perfect fluid energy-momentum tensor
%**   Perfect fluid %%%%%%%%%%%%%%%%%%%%%%%%%%%%%%%%%%%%%%%%%%%%%%%%%%%%%%%%%
\begin{eqnarray}
\tilde T_{\mu\nu} = \rho u_\mu u_\nu + p ( g_{\mu\nu} + u_\mu u_\nu ),
\label{Perfect fluid}
\end{eqnarray}
%%%%%%%%%%%%%%%%%%%%%%%%%%%%%%%%%%%%%%%%%%%%%%%%%%%%%%%%%%%%%%%%%%%
where $u_\mu$ is a unit timelike vector field representing the 4-velocity of the fluid (we simply
choose $u_\mu = (1, 0, 0, 0)$), and $\rho = \rho (t)$ and $p = p (t)$ are respectively the mass-energy
density and the pressure of the fluid, which we assume functions only of the time coordinate $t$.  
With the Robertson-Walker metric (\ref{RW}), the traceless condition $\tilde T^\mu_\mu = 0$ requires
the equation of state
%**   State eq %%%%%%%%%%%%%%%%%%%%%%%%%%%%%%%%%%%%%%%%%%%%%%%%%%%%%%%%%
\begin{eqnarray}
\rho = 3 p.
\label{State eq}
\end{eqnarray}
%%%%%%%%%%%%%%%%%%%%%%%%%%%%%%%%%%%%%%%%%%%%%%%%%%%%%%%%%%%%%%%%%%%

Moreover, using this relation, the conservation law $\nabla_\mu \tilde T^{\mu\nu} = 0$, which is obtained
via Eqs. (\ref{Einstein 1}), can be solved to 
%**   State eq 2 %%%%%%%%%%%%%%%%%%%%%%%%%%%%%%%%%%%%%%%%%%%%%%%%%%%%%%%%%
\begin{eqnarray}
\rho (t) = \frac{A}{a^4 (t)},
\label{State eq 2}
\end{eqnarray}
%%%%%%%%%%%%%%%%%%%%%%%%%%%%%%%%%%%%%%%%%%%%%%%%%%%%%%%%%%%%%%%%%%%
where $A$ is an integration constant which is taken to be positive. Then, Eqs. (\ref{Einstein 1})
reduce to a single differential equation
%**   Cosm-eq %%%%%%%%%%%%%%%%%%%%%%%%%%%%%%%%%%%%%%%%%%%%%%%%%%%%%%%%%
\begin{eqnarray}
\left( \frac{\dot a}{a} \right)^2 + \frac{k}{a^2} - \frac{\Lambda}{3} = \frac{A}{3 a^4}.
\label{Cosm-eq}
\end{eqnarray}
%%%%%%%%%%%%%%%%%%%%%%%%%%%%%%%%%%%%%%%%%%%%%%%%%%%%%%%%%%%%%%%%%%%
Here, in order to have an expanding universe, we assume $\Lambda > 0$. For $k \geq 0$ and 
$A \geq \frac{9 k^2}{4 \Lambda}$, this equation has an interesting solution representing an expanding universe
%**   Cosm-sol %%%%%%%%%%%%%%%%%%%%%%%%%%%%%%%%%%%%%%%%%%%%%%%%%%%%%%%%%
\begin{eqnarray}
a^2 (t) = \left( \frac{4 A}{\Lambda} - \frac{9 k^2}{\Lambda^2} \right)^{\frac{1}{2}} 
\sinh \left( 2 \sqrt{\frac{\Lambda}{3}} t \right) + \frac{3 k}{2 \Lambda}.
\label{Cosm-sol}
\end{eqnarray}
%%%%%%%%%%%%%%%%%%%%%%%%%%%%%%%%%%%%%%%%%%%%%%%%%%%%%%%%%%%%%%%%%%%
Note that this solution describes a universe starting with $a (0) = \sqrt{\frac{3k}{2 \Lambda}}$.

From the solution (\ref{Cosm-sol}), the Hubble parameter reads
%**   H %%%%%%%%%%%%%%%%%%%%%%%%%%%%%%%%%%%%%%%%%%%%%%%%%%%%%%%%%
\begin{eqnarray}
H (t) = \frac{ \sqrt{\frac{\Lambda}{3}} \left( \frac{4 A}{\Lambda} - \frac{9 k^2}{\Lambda^2} \right)^{\frac{1}{2}} 
\cosh \left( 2 \sqrt{\frac{\Lambda}{3}} t \right) }
{ \left( \frac{4 A}{\Lambda} - \frac{9 k^2}{\Lambda^2} \right)^{\frac{1}{2}} 
\sinh \left( 2 \sqrt{\frac{\Lambda}{3}} t \right) + \frac{3 k}{2 \Lambda} }.
\label{H}
\end{eqnarray}
%%%%%%%%%%%%%%%%%%%%%%%%%%%%%%%%%%%%%%%%%%%%%%%%%%%%%%%%%%%%%%%%%%%
Then, for $t \gg 1$, the parameter $\Omega_\Lambda (t)$ takes the form
%**   Omega %%%%%%%%%%%%%%%%%%%%%%%%%%%%%%%%%%%%%%%%%%%%%%%%%%%%%%%%%
\begin{eqnarray}
\Omega_\Lambda (t) \approx \tanh^2 \left( 2 \sqrt{\frac{\Lambda}{3}} t \right)
\rightarrow 1.
\label{Omega}
\end{eqnarray}
%%%%%%%%%%%%%%%%%%%%%%%%%%%%%%%%%%%%%%%%%%%%%%%%%%%%%%%%%%%%%%%%%%%
This result suggests a physical picture such that the universe starts with a radius 
of $\sqrt{\frac{3k}{2 \Lambda}}$ (in case of $k=0$, this radius is zero),
and expands till the time of the electro-weak phase transition and in the process 
the contribution from the cosmological constant gradually becomes dominant.
Of course, in this approach, we have ignored quantum corrections which would
play a critical role particularlly at the beginning of the universe.
Furthermore, the present universe is in the low-energy, Higgs phase where the
spontaneous symmetry breakdown of the gauge symmetry has already occurred,
so our analysis does not provide for any useful information on various problems 
such as the cosmological constant problem.

%%%%%%%%%%%%%%%%%%%%%%%%%%%%%%%%%%%%%%%%%%%%%%%%%%%%%%%%%%%%%%%%%%%%%
%%%%%%%%%%%%%%%%%%%%%%%%%%%%%%   SEC  5    %%%%%%%%%%%%%%%%%%%%%%%%%%
%%%%%%%%%%%%%%%%%%%%%%%%%%%%%%%%%%%%%%%%%%%%%%%%%%%%%%%%%%%%%%%%%%%%%
\section{Conclusion}

In this article, we have constructed a toy model of conformal gravity coupled to the Higgs field,
which can be extended to a more realistic model of conformal gravity coupled to the Standard
Model of elementary particles without any difficulties by incorporating fermions and non-abelian
gauge fields. 

There are two intriguing features in the model at hand. The one feature is that the model has a 
positive Newton's constant, which should be contrasted to the Mannheim's conformal gravity
where the effective Newton's constant must be negative. Perhaps, the positive Newton's constant
makes it possible for our model to account for the observed data of the cosmic microwave background. 

The other feature is that our model provides us with a novel spontaneous symmetry breaking 
mechanism of gauge symmetries. In the conventional quantum field theories, for a complex scalar field $\varphi$ 
the Higgs potential is a priori fixed to be 
%**   Cov-Higgs pot %%%%%%%%%%%%%%%%%%%%%%%%%%%%%%%%%%%%%%%%%%%%%%%%%%%%%%%%%
\begin{eqnarray}
V =  \frac{\lambda}{4} ( \varphi^\dagger \varphi )^2 - m^2 \varphi^\dagger \varphi,
\label{Cov-Higgs pot}
\end{eqnarray}
%%%%%%%%%%%%%%%%%%%%%%%%%%%%%%%%%%%%%%%%%%%%%%%%%%%%%%%%%%%%%%%%%%%
where the sign in front of the mass term is negative, that is, tachyonic mass. Even if this Higgs potential
explains the value of the Higgs mass, it is quite difficult to understand its origin. In our model,
it is the local conformal symmetry and its gauge-fixing that determines the form of the Higgs potential,
so the origin is very clear. Moreover, the gauge-fixing condition is not arbitrary at all, which is
uniquely fixed by requiring that the correct Newton's constant is reproduced in our model.

There remain many of unsolved issues within the framework of the present formalism. For instance, our
analysis is purely classical, so we should investigate quantum corrections. In this context, let us
note that the renormalizability of our model makes it possible to calculate quantum corrections by perturbation
theory. Another interesting problem is to investigate classical solutions. We wish to return these
issues in near future.

%%%%%%%%%%%%%%%%%%%%%%%%%%%%%%%%%%%%%%%%%%%%%%%%%%%%%%%%%%%%%%%%%%
%%%%%%%%%%%%%%%%%%%%%%%% Acknowledgements %%%%%%%%%%%%%%%%%%%%%%%%%%%%%
%%%%%%%%%%%%%%%%%%%%%%%%%%%%%%%%%%%%%%%%%%%%%%%%%%%%%%%%%%%%%%%%%%
\begin{flushleft}
{\bf Acknowledgements}
\end{flushleft}
This work is supported in part by the Grant-in-Aid for Scientific 
Research (C) No. 25400262 from the Japan Ministry of Education, Culture, 
Sports, Science and Technology.

%%%%%%%%%%%%%%%%%%%%%%% reference %%%%%%%%%%%%%%%%%%%%%%%%%%%%%%%
%%%%%%%%%%%%%%%%%%%%%%%%%%%%%%%%%%%%%%%%%%%%%%%%%%%%%%%%%%%%%%%%%%

\end{document}